\documentclass{ws-procs975x65}

\def\beq{\begin{equation}}
\def\eeq{\end{equation}}
\def\nn{\nonumber}
\def\x{\mathbf{x}}
\def\k{\mathbf{k}}
\def\rnd{\partial}
\def\A{\mathrm{A}}
\def\inv{\mathrm{in}}

\begin{document}

\title{Effect of Schwinger pair production on the evolution of the Hubble constant in de~Sitter spacetime}
\author{Ehsan Bavarsad$^*$}

\address{Department of Physics, University of Kashan, Kashan, 8731753153, Iran\\
$^*$E-mail: bavarsad@kashanu.ac.ir}

\author{Sang Pyo Kim}

\address{Department of Physics, Kunsan National University, Kunsan, 54150, Korea\\
Institute of Theoretical Physics, Chinese Academy of Sciences, Beijing, 100190, China\\
E-mail: sangkim@kunsan.ac.kr}

\author{Cl\'{e}ment Stahl}

\address{Instituto de F\'{\i}sica, Pontificia Universidad Cat\'{o}lica de Valpara\'{\i}so, Casilla, Valpara\'{\i}so, 4950, Chile\\
E-mail: clement.stahl@pucv.cl}

\author{She-Sheng~Xue}

\address{ICRANet, Piazzale della Repubblica, Pescara,  10, 65122 ,Italy\\
Dipartimento di Fisica, Universita di Roma ``La Sapienza'', Piazzale Aldo Moro, Rome, 5, 00185, Italy\\
E-mail: xue@icra.it}

\begin{abstract}
In this proceeding we consider a massive charged scalar field in a uniform electric field background in a de~Sitter spacetime (dS).
We compute the in-vacuum expectation value of the trace of the energy-momentum tensor for the created Schwinger pairs, and using adiabatic subtraction scheme the trace is regularized. The effect of the Schwinger pair creation on the evolution of the Hubble constant is investigated.
We find that the production of the semiclassical pairs leads to a decay of the Hubble constant.
Whereas, the production of a light scalar field in the weak electric field regime leads to a superacceleration phenomenon.
\end{abstract}
\bodymatter
\section{\label{sec:intro}Introduction}
In a flat spacetime the phenomenon of the pair creation in a strong electric field background is referred to as the Schwinger effect.
\cite{Schwinger:1951nm} The presence of the strong electromagnetic and gravitational fields in the early Universe motivates the study of the Schwinger
effect in dS.\cite{Martin:2007bw} The Schwinger effect and the induced current of the created pairs were investigated in a uniform electric field for various dimensions of the de Sitter spacetime in Refs.~\refcite{Garriga:1994bm,Frob:2014zka,Kobayashi:2014zza,Bavarsad:2016cxh,Kim:2008xv}.
Recently, the effect of a conserved flux magnetic field on the Schwinger effect in dS has been investigated in Ref.~\refcite{Bavarsad:2017oyv}, see also Ref.~\refcite{Bavarsad:2018lvn}.
Without considering an electromagnetic field, the energy-momentum tensor of the created scalar particles in the gravitational background field of dS has been investigated in Refs.~\refcite{Mottola:1984ar,Parker:1974qw,Fulling:1974zr,Dowker:1975tf,Habib:1999cs,Markkanen:2016aes,Markkanen:2017abw,Gibbons:1977mu,Zhang:2019urk}.
The results of Refs.~\refcite{Mottola:1984ar,Markkanen:2016aes,Markkanen:2017abw} show that the Hubble constant of dS decays due to the particle creation.
Furthermore, the gravitational backreaction effects of the quantum fluctuations may lead to a superacceleration phase, where the Hubble constant
increase.\cite{Markkanen:2016aes}
Using a semiclassical approach the energy-momentum tensor of the created Schwinger scalars in a dS has been computed in Refs.~\refcite{Bavarsad:2016cxh,Bavarsad:2017wbe}, which showed that the Hubble constant decays.
The trace of the induced energy-momentum tensor of the Schwinger scalars in a three dimensional dS has been investigated
in  Ref.~\refcite{Bavarsad:ijpr}, and the authors found that the creation of the semiclassical pairs may lead to an increase of the Hubble constant.
With the aim of developing the renormalization theory in the curved spacetime, recently the
Schwinger effect and the conformal anomaly for both of the scalar and spinor de~Sitter QED cases have been investigated in
Ref.~\refcite{Ferreiro:2018qdi}.
With the aim of completing the work started in Ref.~\refcite{Bavarsad:2016cxh} and with cosmological applications in mind, in this proceeding we compute
the trace of the energy-momentum tensor for the Schwinger scalars created in a uniform electric field background in a four dimensional dS.
The proceeding is organised as follows: in Sec.~\ref{sec:pre} the preliminary explanation of our model is introduced.
We compute the regularized trace of the induced energy-momentum tensor in Sec.~\ref{sec:emt}. In Sec.~\ref{sec:concl} we give some conclusions.
\section{\label{sec:pre}Preliminaries}
We begin our study by considering the scalar QED action in a four dimensional dS as
\begin{equation}\label{action}
S=\int{d^4}x\sqrt{|g|}\Big\{g^{\mu\nu}\big(\rnd_{\mu}+ieA_{\mu}\big)\varphi\big(\rnd_{\nu}-ieA_{\nu}\big)
\varphi^{\ast}-\big(m^{2}+\xi R\big)\varphi\varphi^{\ast}-\frac{1}{4}F_{\mu\nu}F^{\mu\nu}\Big\},
\end{equation}
where $\varphi(x)$ is a complex scalar field with mass $m$ and electric charge $e$ which is coupled to an electromagnetic vector potential background $A_{\mu}(x)$. The parameter $\xi$ is a dimensionless nonminimal coupling of the scalar field to the Ricci scalar $R$, and in this proceeding from now on
we set $\xi=1/6$. The $|g|$ denotes the absolute value of the metric determinant.
In Secs.~\ref{sec:pre} and~\ref{sec:emt}, we assume that the gravitational and electromagnetic fields are not affected by the pair creation.
This assumption on the gravitational field does not hold when we discuses backreaction effects in Sec.~\ref{sec:concl}.
We consider the Poincar\'{e} patch of dS, whose metric reads
\begin{align}\label{metric}
ds^{2} &= \Omega^{2}(\tau)\big(d\tau^{2}-d\x^{2}\big), && \Omega(\tau)=-\frac{1}{\tau H}, && \tau \in(-\infty,0), && \x \in \mathbb{R}^{3},
\end{align}
where $H$ is the Hubble constant and $\tau$ is the conformal time which relates to the cosmological proper time $t$ as
\begin{equation}
\tau=-\frac{1}{H}e^{-Ht}. \label{time}
\end{equation}
Having a uniform electric field with the constant energy density in the metric background given by Eq.~(\ref{metric}), we choose the electromagnetic
vector potential in the gauge
\begin{equation}
A_{\mu}(\tau)=-\frac{E}{H^{2}\tau}\delta_{\mu}^{1}, \label{vector}
\end{equation}
where $E$ is a constant. The the Klein-Gordon equation for the scalar field reads from the action (\ref{action}). We impose that the mode functions of the Klein-Gordon equation have the asymptotic behavior similar to those mode functions in the Minkowski spacetime at the early times,
$\tau\rightarrow-\infty$. Then, these mode functions that describe the Hadamard\cite{Frob:2014zka} in vacuum state are given by\cite{Kobayashi:2014zza,Bavarsad:2016cxh}
\begin{eqnarray}
\label{uin}
U_{\inv\k}(x)&=&(2k)^{-\frac{1}{2}}e^{\frac{i\pi\kappa}{2}}\Omega^{-1}(\tau)e^{+i\k\cdot\x}W_{\kappa,\gamma}\big(2ik\tau\big), \\
\label{vin}
V_{\inv\k}(x)&=&(2k)^{-\frac{1}{2}}e^{-\frac{i\pi\kappa}{2}}\Omega^{-1}(\tau)e^{-i\k\cdot\x}W_{\kappa,-\gamma}\big(-2ik\tau\big),
\end{eqnarray}
where $U_{\inv\k}$ and $V_{\inv\k}$ are positive and negative frequency mode functions, respectively. The function $W$ is the Whittaker function,
and the variables are defined as
\begin{align}\label{zpm}
k&:=\big|\k\big|, &\mu &:=\frac{m}{H}, & \lambda&:=-\frac{eE}{H^{2}}, & r:=&\frac{k_{x}}{k}, & \kappa&:=-i\lambda r, &
\gamma^{2}&:=\frac{1}{4}-\mu^{2}-\lambda^{2}.
\end{align}
Then we can expand the scalar field operator $\phi(x)$ as
\begin{equation}\label{phiin}
\phi(x)=\int\frac{d^{3}k}{(2\pi)^{3}}\Big[U_{\inv\k}(x)a_{\inv\k}+V_{\inv\k}(x)b^{\dag}_{\inv\k}\Big],
\end{equation}
where $a_{\inv\k}$ and $b^{\dag}_{\inv\k}$ are the annihilation and creation operators for particle and antiparticles with comoving momentum $\k$, respectively, which satisfy the commutation relations. Then, the in vacuum state is defined by
\begin{align}
a_{\inv\k} |\inv\rangle=b_{\inv\k} |\inv\rangle=0, && \forall\,\k. \label{vacin}
\end{align}
\section{\label{sec:emt}Trace of the Induced Energy-Momentum Tensor}
\begin{figure}[t]
\begin{center}
\includegraphics[width=4.5in]{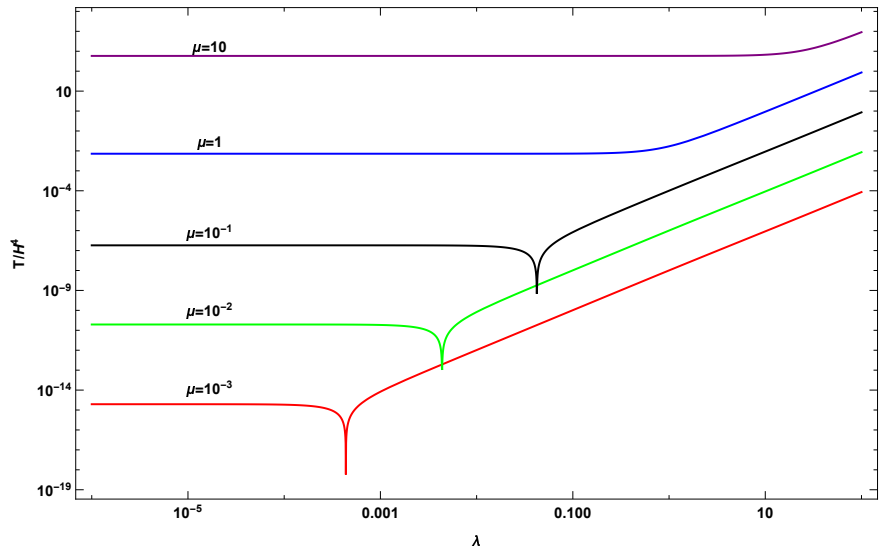}
\end{center}
\caption{The normalized trace $T/H^{4}$ is plotted as functions of the normalized electric field $\lambda=|-eE|/H^{2}$, for different values of the normalized scalar field mass $\mu=m/H$.}
\label{fig:1}
\end{figure}

The induced energy-momentum tensor of the created Schwinger pairs is necessary to study evolution of the de~Sitter spacetime.
We will compute all of the induced energy-momentum tensor's components in a future work. However, in this proceeding we assume that the created Schwinger
pairs take the form of a perfect fluid with the \textit{vacuum} equation of state. Consequently, the energy-momentum tensor $T_{\mu\nu}$ is related to
its trace $T$ as
\begin{equation}
T_{\mu\nu}=\frac{1}{4}Tg_{\mu\nu}. \label{state}
\end{equation}
Hence, it is sufficient to compute the trace of the induced energy-momentum tensor which is defined by the variation of the action~(\ref{action}) with respect to the variation of the metric $\delta g_{\mu\nu}$ as
\begin{equation}
T:=-\frac{2}{\sqrt{|g|}}g_{\mu\nu}\frac{\delta S}{\delta g_{\mu\nu}}.  \label{variation}
\end{equation}
In the action (\ref{action}) the Maxwell term, $F_{\mu\nu}F^{\mu\nu}$, is the pure electromagnetic part of the action which is constant, and according to
our assumption is not affected by the Schwinger pair creation, hence to obtain the trace of the induced energy-momentum tensor of the Schwinger pairs we
do not consider this term. We then obtain
\begin{equation}
T = 2m^{2}\varphi\varphi^{\ast}. \label{temt}
\end{equation}
Substituting the scalar field operator~(\ref{phiin}) into the expression~(\ref{temt}) and using Eq.~(\ref{vacin}) leads to
\begin{equation}
\langle \inv |T| \inv \rangle = 2m^{2}\int\frac{d^{3}k}{(2\pi)^{3}}\Big|U_{\inv\k}\big(x)\Big|^{2}. \label{vev}
\end{equation}
Equation~(\ref{vev}), in terms of a dimensionless integral variable $p=-k\tau$ can be written as
\begin{eqnarray}
\langle \inv| T| \inv \rangle = \frac{H^{4}\mu^{2}}{4\pi^{2}} \int_{-1}^{1}dr
e^{\lambda\pi r}\int_{0}^{\Lambda}dp\,p\Big|W_{-i\lambda r,\gamma}\big(-2ip\big)\Big|^{2}, \label{intt}
\end{eqnarray}
where the dimensionless momentum cutoff $\Lambda$ is defined to regularize the ultraviolet divergency. Following the integration procedure introduced in Refs.~\refcite{Frob:2014zka,Kobayashi:2014zza} we obtain the unregularized expression for the trace of the induced energy-momentum tensor
\begin{align}
&\langle \inv| T| \inv \rangle = \frac{H^{4}\mu^{2}}{16\pi^{2}}\bigg\{4\Lambda^{2}-4\mu^{2}\log(2\Lambda)+2\mu^{2}-\frac{8\lambda^2}{3}
+2+2i\pi\mu^{2}-\frac{6\gamma}{\pi}\csc(2\pi\gamma)\nn\\
&\times\Big(\cosh(2\pi\lambda)-\frac{1}{2\pi\lambda}\sinh(2\pi\lambda)\Big)
+i\csc(2\pi\gamma)\int_{-1}^{1}dr\big(3r^{2}\lambda^{2}-\lambda^{2}-\mu^{2}\big) \nn\\
&\times\bigg[\big(e^{2\pi\lambda r}+e^{2\pi i\gamma}\big)\psi\Big(\frac{1}{2}+i\lambda r+\gamma\Big)-\big(e^{2\pi\lambda r}+e^{-2\pi i\gamma}\big)\psi\Big(\frac{1}{2}+i\lambda r-\gamma\Big)\bigg]\bigg\}, \label{unreg}
\end{align}
where $\psi$ is the digamma function. In order to remove the ultraviolet divergences from the expression (\ref{unreg}) we apply the adiabatic subtraction scheme as introduced in Ref.~\refcite{book}. However, a new condition for renormalization the vacuum expectation value of the quantities in the context of de~Sitter QED has been introduced in Ref.~\refcite{Hayashinaka:2018amz}. The positive frequency mode function of adiabatic zeroth order is given by
\begin{equation}
U_{\A}(x)=\Omega^{-1}(\tau)\Big(2\omega(\tau)\Big)^{-\frac{1}{2}}\exp\Big(i\k.\x-i\int \omega(\tau)d\tau\Big), \label{WKB}
\end{equation}
where the time dependent frequency is
\begin{equation}
\omega(\tau)=H\Omega(\tau)\sqrt{k^{2}\tau^{2}+2\lambda rk\tau+\lambda^{2}+\mu^{2}}. \label{frequency}
\end{equation}
Using the mode function (\ref{WKB}) the zeroth order of the adiabatic expansion of the trace of the energy-momentum tensor is obtained
\begin{equation}
T_{\A}=\frac{H^{4}\mu^{2}}{16\pi^{2}}\bigg(4\Lambda^{2}-4\mu^{2}\log(2\Lambda)+2\mu^{2}-\frac{8\lambda^2}{3} \bigg). \label{TA}
\end{equation}
The adiabatic regularized trace of the induced energy-momentum tensor is given by subtracting the counterterm (\ref{TA}) from the unregularized
expression (\ref{unreg}). We then obtain the regularized trace as
\begin{align}
T&=\langle \inv| T| \inv \rangle - T_{\A}=\frac{H^{4}\mu^{2}}{16\pi^{2}}\bigg\{2+2i\pi\mu^{2}-\frac{6\gamma}{\pi}\csc(2\pi\gamma)
\Big(\cosh(2\pi\lambda)\nn\\
&-\frac{1}{2\pi\lambda}\sinh(2\pi\lambda)\Big)+i\csc(2\pi\gamma)\int_{-1}^{1}dr\big(3r^{2}\lambda^{2}-\lambda^{2}-\mu^{2}\big)
\bigg[\big(e^{2\pi\lambda r}+e^{2\pi i\gamma}\big) \nn\\
&\times\psi\Big(\frac{1}{2}+i\lambda r+\gamma\Big)-\big(e^{2\pi\lambda r}
+e^{-2\pi i\gamma}\big)\psi\Big(\frac{1}{2}+i\lambda r-\gamma\Big)\bigg]\bigg\}. \label{reg}
\end{align}
In Fig.~\ref{fig:1} the regularized trace~(\ref{reg}) is plotted as function of the electric field for different values of the scalar field mass.
A numerical investigation shows that for a massive scalar field $\mu\gtrsim 1$, the sign of the trace is positive. However, for a light scalar field
$\mu\lesssim 1$, the trace vanishes at $\lambda=L$. In the domain $\lambda<L$ the sign of the trace is negative, whereas in the domain $\lambda>L$ the
sign of the trace is positive. Further numerical investigations illustrate that $L\simeq 0.4\mu$. Therefore in the semiclassical regime
$\lambda^{2}+\mu^{2}\gg 1$ the sign of the trace is positive, whereas in the infrared regime $\lambda^{2}+\mu^{2}\ll 1$ the sign of the trace is negative.
\section{\label{sec:concl}Conclusions}
These results for the trace of the energy-momentum tensor would be important for discussing the gravitational backreaction effect of the Schwinger pair creation. As a consequence of considering the \textit{vacuum} equation of state for the created Schwinger pairs, see Eq.~(\ref{state}), the Einstein equation leads to the time evolution equation for the Hubble constant as \cite{Mottola:1984ar}
\begin{equation}
\frac{dH}{dt}=-\frac{\pi T}{3M_{\mathrm{P}}^{2}},\label{evolution}
\end{equation}
where $t$ is the proper cosmological time, see Eq.~(\ref{time}), and $M_{\mathrm{P}}$ is the Planck mass. Considering the results of Sec.~\ref{sec:emt}
for the trace~(\ref{reg}) and Eq.~(\ref{evolution}), we conclude that in the semiclassical regime $\lambda^{2}+\mu^{2}\gg 1$, that the sign of the trace
is positive the Hubble constant decays $dH/dt<0$. This result is in agreement with the results obtained in Ref.~\refcite{Bavarsad:2016cxh} for the decay
of the Hubble constant due to the semiclassical Schwinger pair creation.
We find that in the infrared regime $\lambda^{2}+\mu^{2}\ll 1$, that the sign of the trace is negative, Eq.~(\ref{evolution}) implies that $dH/dt>0$.
In Ref.~\refcite{Markkanen:2016aes} the authors found that the gravitational backreaction effects of the quantum fluctuations may lead to a similar
behavior of the Hubble constant, i.e., a period of superacceleration with $dH/dt>0$.
\section*{Acknowledgments}
E.~B. is supported by the University of Kashan.

\end{document}